\documentclass[preprint,12pt]{elsarticle}

\usepackage{amsmath,amssymb}
\usepackage{siunitx}
\usepackage{booktabs}
\usepackage{hyperref}
\usepackage{bm}
\usepackage{empheq}
\usepackage{enumitem}
\usepackage{xcolor}

\usepackage{fancyhdr}
\usepackage{xurl} 

\journal{Optics Communications}

\newcommand{\vr}{\mathbf r}
\newcommand{\vomega}{\hat{\boldsymbol\omega}}

\fancypagestyle{plain}{%
  \fancyhf{}
  \fancyhead[C]{%
    \footnotesize
    {\color{red}\textbf{Author's preprint.}} A peer-reviewed version of this article was published in
    \textbf{\emph{Optics Communications}} \textbf{608} (2026), Article 133016.
    Version of Record:
    \href{https://doi.org/10.1016/j.optcom.2026.133016}{DOI: 10.1016/j.optcom.2026.133016}%
  }
  \fancyfoot[C]{\thepage}
}

\begin{document}

\setlength{\headheight}{28pt} 

\begin{frontmatter}

\title{Instrument-limited pixel-level SNR bounds from optical throughput\tnoteref{pubnote}}

\author[1]{Jan Sova}
\author[1]{Marie Kola\v{r}\'{i}kov\'{a}}
\address[1]{Czech Technical University in Prague, Faculty of Mechanical Engineering,
Department of Manufacturing Technology, Technick\'{a} 4, 16607 Prague, Czech Republic}

\tnotetext[pubnote]{\textcolor{red}{\textbf{Author's preprint.}} A peer-reviewed version of this article was published in \textbf{\emph{Optics Communications}} \textbf{608} (2026), Article 133016. Version of Record: \href{https://doi.org/10.1016/j.optcom.2026.133016}{DOI: 10.1016/j.optcom.2026.133016}.}

\begin{abstract}
The radiometric integral is the fundamental radiance--to--flux relation in imaging, whereas \'etendue is typically used as a compact system-level descriptor.
For quantitative imaging and calibration, however, the operative mapping must be explicit at the level of individual detector pixels, including pixel acceptance and field-dependent pupil visibility.

This work packages the pixel-restricted radiometric integral into a reusable geometric throughput factor by defining a per-pixel optogeometric (optical-throughput) factor $F_{\mathrm{opg},i}$ (units \si{m^2.sr}) such that, under weak radiance variation,
$\Phi_i \approx L_i\,F_{\mathrm{opg},i}$.
Making throughput explicit at the pixel scale yields an optics-delivered photon budget in which the incident photon count at the detector, $N_{\mathrm{inc},i}$ (before quantum efficiency), scales linearly with geometry:
$N_{\mathrm{inc},i}\propto F_{\mathrm{opg},i}$ for a given scene radiance distribution and fixed acquisition settings (bandwidth, integration time, and optical transmission).
The corresponding optics-delivered (pre-detection) shot-noise ceiling is set by the incident photon count
$N_{\mathrm{inc},i}$, with $\mathrm{SNR}_{\mathrm{inc},i}\le \sqrt{N_{\mathrm{inc},i}}\propto \sqrt{F_{\mathrm{opg},i}}$,
while in photoelectron units one has
$\mathrm{SNR}_i \le \sqrt{N_{\mathrm{ph},i}}=\sqrt{\eta(\bar\nu)\,N_{\mathrm{inc},i}}\propto \sqrt{F_{\mathrm{opg},i}}$,
where $N_{\mathrm{ph},i}$ is the detected photoelectron count and $\eta(\bar\nu)$ is the (narrowband) quantum efficiency;
additional detector/electronics noise sources (e.g.\ dark current and read noise) can only reduce the achieved SNR below these shot-noise limits.

To compare throughput across wavelength bands, a throughput-normalized pixel-level phase-space proxy
$\mathcal{M}_i(\lambda)\equiv F_{\mathrm{opg},i}/\lambda^2$ is introduced.
In the paraxial unvignetted baseline, $F_{\mathrm{opg},i}$ reduces to a compact design expression, recovering the familiar scaling $\mathrm{SNR}_i \propto a_{\mathrm{pp},i}/(f\#\,|M|)$ as a corollary.
Overall, the results remain within standard radiometry but elevate pixel-resolved throughput to a first-class quantity, providing an explicit end-to-end link \emph{pixel radiometry $\rightarrow$ optical throughput $\rightarrow$ photon budget $\rightarrow$ SNR} for thermography, remote sensing, and quantitative imaging.
\end{abstract}

\end{frontmatter}

\section{Introduction}
\label{sec:intro}

The relation between scene radiance and collected flux is a cornerstone of radiometry and imaging
\cite{grant2011field,boyd1983radiometry,howell2020thermal}.
In idealized passive optical mappings (no absorption, scattering, or vignetting),
the area--solid-angle product (\'etendue) is invariant and underpins imaging design and performance analysis
\cite{welford1978high,born1999principles,goodman2005fourier}.
For quantitative imaging, metrology, and calibration workflows, however, the operative statements are often required at the level of
\emph{individual detector pixels}: how much scene radiance is mapped into the radiant flux collected by a single detector element,
how many throughput-normalized spatial degrees of freedom that pixel admits, and how these features constrain photon-limited signal-to-noise ratio (SNR).
A practical complication is that system-level \'etendue alone does not provide a reusable \emph{pixel-resolved} radiance--to--flux mapping once pixel acceptance,
field-dependent pupil visibility (vignetting/apodization), and other non-idealities are made explicit.
The present work isolates this mapping by defining a per-pixel optogeometric throughput $F_{\mathrm{opg},i}$ directly from the radiometric integral, where index $i$ labels detector pixels.
This definition makes explicit the end-to-end chain
\emph{pixel radiometry $\rightarrow$ optical throughput $\rightarrow$ photon budget $\rightarrow$ shot-noise SNR bound}.
Because the optical dependence enters only through $F_{\mathrm{opg},i}$, the formulation remains portable across aperture changes, vignetting conditions, and pixel-level calibration workflows. 

Pixel-level formulations that treat vignetting and pupil visibility in a phase-space (light-field) setting have been discussed, e.g.\ in the context of light-field camera vignetting models \cite{MignardDebiseIhrke2019}.

This explicit optics-delivered photon budget is particularly relevant in applied thermography and related quantitative imaging, where \emph{apparent} SNR improvements are often reported after
post-processing steps such as non-uniformity correction and compensation of non-uniform excitation/illumination
\cite{chung2024LIT,giron2013nonuniformheating,boutemedjet2016nuc}.
Because these procedures operate at the detector-signal level, the underlying per-pixel photon budget set by optical throughput is typically kept implicit,
making it difficult to separate genuine throughput-limited gains from sensor/electronics corrections in a physically interpretable way.

\subsection{Motivation and scope}
\label{sec:motivation_scope}

In practice, radiometric performance is often treated in layered terms, i.e., as a stack of partially decoupled models for calibration (radiance-to-signal), optics (vignetting/aberrations), and noise/SNR, which are frequently reported in different conventions.

Accordingly, calibration-oriented forward models address the radiance-to-digital-number transfer, while optical aberrations and image-formation effects may be handled in separate correction steps
\cite{Ngo2019RadiometricCalibrationVNREDSat,Tansock2015NISTHB157}.
Conversely, SNR formulations and noise models are frequently presented separately from calibration-oriented
radiometric modeling, and conventions (e.g.\ exposure-referred vs.\ radiance-referred definitions) may differ across sources,
complicating direct comparison and interpretation \cite{Gnanasambandam2022ExposureReferredSNR}.
In remote sensing, radiance-referred SNR figures are commonly reported; however, their physical interpretation at the detector-pixel level is not unique unless tied to an explicit per-pixel photon budget (bandwidth, integration time, processing level) and the optical throughput that maps scene radiance into incident flux/photoelectrons \cite{Kudela2024ExpandedSNR}.

Although the motivating applications include thermography and remote sensing, the formulation is not infrared-specific.
The optogeometric factor $F_{\mathrm{opg},i}$ defined here is a purely geometric, pixel-level throughput that applies to any imaging detector
(CMOS/CCD in the visible and near-IR, SWIR, and thermal IR), because it arises directly from the radiometric integral.
Across wavelength bands and detector technologies, the quantities that change are the spectral radiance $L_{\nu,i}(\nu)$, the optical transmission $\tau_{\mathrm{opt}}(\nu)$, and the quantum efficiency $\eta(\nu)$,
whereas the geometric radiance-to-flux mapping encoded by $F_{\mathrm{opg},i}$ remains the same (for a fixed optical geometry and acceptance model).
Consequently, for a given scene and fixed detection settings, the geometric scaling $\mathrm{SNR}_i\propto\sqrt{F_{\mathrm{opg},i}}$
provides a compact design- and calibration-oriented link between optics (aperture, vignetting, pixel iFOV) and pixel-level photon statistics.

\subsection{Contributions}
The following results are obtained.

\begin{enumerate}[leftmargin=*, itemsep=0.4em]

\item A per-pixel optogeometric factor $F_{\mathrm{opg},i}$ is defined directly from the pixel-restricted radiometric integral,
including a general pupil-visibility weighting $T_i(\mathbf r,\hat{\boldsymbol\omega})\in[0,1]$ that captures geometric pupil visibility
(e.g.\ vignetting and apodization).

\item At wavelength $\lambda$, normalization by $\lambda^2$ defines a throughput-normalized per-pixel degree-of-freedom proxy per polarization,
$\mathcal{M}_i(\lambda)\equiv F_{\mathrm{opg},i}/\lambda^2$.

\item A first-principles photon budget yields an explicit shot-noise bound
$\mathrm{SNR}_i \propto \sqrt{F_{\mathrm{opg},i}}$ for a given scene and fixed detection settings.
A key contribution is the explicit end-to-end link \emph{pixel radiometry $\rightarrow$ optical throughput $\rightarrow$ pixel SNR}:
starting from the pixel-restricted radiometric integral (radiance-to-flux), the geometric throughput factor $F_{\mathrm{opg},i}$
enters the photon budget linearly ($N_{\mathrm{inc},i}\propto F_{\mathrm{opg},i}$) and therefore constrains the photon-limited SNR
as $\mathrm{SNR}_i \propto \sqrt{F_{\mathrm{opg},i}}$.
Because the optical dependence enters only through $F_{\mathrm{opg},i}$, the same bound applies with any appropriate evaluation of
$F_{\mathrm{opg},i}$ (paraxial, finite-cone, vignetted, or numerical).
A finite-cone correction provides one such refinement, e.g.\ Eq.~\eqref{eq:fopg_correction_factor},
quantifying deviations from the small-angle $1/(f\#)^2$ baseline for fast optics.
In the paraxial unvignetted baseline, the result recovers the familiar design scaling
$\mathrm{SNR}_i \propto a_{\mathrm{pp},i}/(f\#\,|M|)$.

\item The separation between geometric throughput (encoded in $F_{\mathrm{opg},i}$) and detector/electronics response non-uniformity
(typically handled by pixel-wise gain/offset correction) is clarified, motivating the usefulness of an explicit per-pixel radiance-to-flux mapping
in calibration and non-uniformity correction (NUC) workflows.

\end{enumerate}

The thermography forward measurement model for oblique-view infrared measurements in \cite{hs2025} employs a \textbf{per-pixel} geometric multiplier. Building on that pixel-level perspective, the present work derives directly from the radiometric integral an explicit detector-agnostic per-pixel radiance-to-flux (optical-throughput) factor $F_{\mathrm{opg},i}$ with clear units and limiting forms, making such geometric prefactors transparent and reusable beyond the specific assumptions of \cite{hs2025}.

System-level thermography measurement equations have been used for a long time \cite{vollmer2010,minkina2009infrared,kaplan2007,ForejtovaAutomotive}.
In practice, however, calibration is often done pixel by pixel: non-uniformity/flat-field correction estimates a gain and an offset for each detector element to suppress fixed-pattern artifacts \cite{Cukor2019,perry_nuc,harris_nuc}.
These procedures typically operate at the detector-signal level and keep the optics-delivered per-pixel photon budget implicit.
An explicit geometric throughput factor $F_{\mathrm{opg},i}$ makes the separation clearer: gain/offset terms describe pixel-wise detector/electronics response non-uniformity, while $F_{\mathrm{opg},i}$ captures the geometric component of the radiance-to-photon mapping that sets the per-pixel photon budget and therefore the best achievable shot-noise-limited SNR.

While two compact baselines are used here (the paraxial unvignetted form in Eq.~\eqref{eq:fopg_paraxial} and the finite-cone refinement in Eqs.~\eqref{eq:fopg_finite_cone}--\eqref{eq:fopg_correction_factor}),
$F_{\mathrm{opg},i}$ is fundamentally defined by the radiometric integral and may be evaluated instrument-specifically (analytically or numerically) to capture the actual optics and pixel acceptance beyond these idealized cases.

\section{Optogeometric factor}
\label{sec:opg}

This section defines the optogeometric factor $F_{\mathrm{opg},i}$ as a pixel-level analogue of geometric optical throughput and \'etendue.
The standard radiometric integral is specialized to the footprint and acceptance set of a single detector element, yielding a compact geometric factor that maps radiance to pixel-collected radiant flux.
A paraxial unvignetted baseline then connects the general definition to familiar design scalings involving pixel pitch, $f$--number, and magnification, and an optional finite-cone (exact-cone) refinement quantifies deviations from the small-angle baseline for fast optics.

The assumptions used in reducing the general definition of $F_{\mathrm{opg},i}$ to the paraxial baseline (and the finite-cone refinement) are stated explicitly.

\subsection{Radiometric integral at the pixel level}
\label{sec:radiometric}

In this section, $L(\mathbf r,\hat{\boldsymbol\omega})$ denotes band-integrated radiance (units \si{W.m^{-2}.sr^{-1}}) at surface position $\mathbf r$ in direction $\hat{\boldsymbol\omega}$, where directions $\hat{\boldsymbol\omega}$ are defined locally at $\mathbf r$ and the pupil-reachable subset may vary across the footprint.
The spectral form required for photon counting is introduced in Sec.~\ref{sec:photon_budget}.

Let $A_{\mathrm{pix},i}$ denote the object-space footprint mapped onto detector pixel $i$, and let $\Omega_i^+(\mathbf r)$ be the set of directions from $\mathbf r\in A_{\mathrm{pix},i}$ that reach the entrance pupil (i.e.\ the pupil-reachable subset of the forward hemisphere; the pixel acceptance set in object space).
With $\mathbf n(\mathbf r)$ the outward unit normal and $\hat{\boldsymbol\omega}$ a unit propagation direction (locally parameterized by $(\theta,\psi)$), the radiant flux collected by pixel $i$ is \cite{grant2011field,boyd1983radiometry,mccluney1994radiometry}

\begin{equation}
\Phi_i
=
\iint_{A_{\mathrm{pix},i}}
\iint_{\Omega_i^+(\mathbf r)}
L(\mathbf r,\hat{\boldsymbol\omega})\,
T_i(\mathbf r,\hat{\boldsymbol\omega})\,
\bigl(\mathbf n(\mathbf r)\!\cdot\!\hat{\boldsymbol\omega}\bigr)\,
\mathrm d\Omega\,\mathrm dA,
\label{eq:radiometric_integral}
\end{equation}

where

\begin{itemize}[leftmargin=*, itemsep=0.25em]
\item $\Phi_i$ [$\si{W}$] is the radiant flux collected by pixel $i$,

\item $A_{\mathrm{pix},i}$ [$\si{m^2}$] is the object-space footprint imaged onto pixel $i$,

\item $\mathbf r\in A_{\mathrm{pix},i}$ denotes a surface point on the object (scene) plane,

\item $\Omega_i^+(\mathbf r)$ is the set of directions from $\mathbf r$ that reach the entrance pupil
(i.e.\ the pupil-reachable subset of the forward hemisphere; the pixel acceptance set in object space),

\item $\hat{\boldsymbol\omega}$ is a unit propagation direction in $\Omega_i^+(\mathbf r)$ (i.e.\ a pupil-reachable direction from $\mathbf r$); for brevity, it is used as the angular integration variable, and its explicit local parametrization (when needed) is written as $\hat{\boldsymbol\omega}(\mathbf r;\theta,\psi)$,

\item $L(\mathbf r,\hat{\boldsymbol\omega})$ [$\si{W.m^{-2}.sr^{-1}}$] is the band-integrated radiance at position $\mathbf r$
in direction $\hat{\boldsymbol\omega}$ (spectral radiance is introduced in Sec.~\ref{sec:photon_budget}),

\item $T_i(\mathbf r,\hat{\boldsymbol\omega})\in[0,1]$ [$-$] is a dimensionless throughput weighting that captures
\emph{geometric pupil visibility} for pixel $i$ (e.g.\ vignetting and pupil apodization) along the path from
$(\mathbf r,\hat{\boldsymbol\omega})$ to pixel $i$; the ideal unvignetted baseline corresponds to $T_i\equiv 1$,

\item $\mathbf n(\mathbf r)$ is the outward unit normal at $\mathbf r$, and
$\mathbf n(\mathbf r)\!\cdot\!\hat{\boldsymbol\omega}$ is the foreshortening factor (directional cosine);
directions with $\mathbf n(\mathbf r)\!\cdot\!\hat{\boldsymbol\omega}\le 0$ do not contribute,

\item $\mathrm dA$ [$\si{m^2}$] is the differential area element on the object surface, and
$\mathrm d\Omega$ [$\si{sr}$] is the differential solid-angle element with
$\mathrm d\Omega = \sin\theta\,\mathrm d\theta\,\mathrm d\psi$.
\end{itemize}

The weighting $T_i(\mathbf r,\hat{\boldsymbol\omega})$ may be interpreted as a pupil-visibility (vignetting or apodization) term; related phase-space treatments of vignetting appear in light-field camera modeling \cite{MignardDebiseIhrke2019}.

Since $0\le T_i(\mathbf r,\hat{\boldsymbol\omega})\le 1$ pointwise over the integration domain,
an upper bound is obtained from the ideal baseline:
\begin{equation*}
\Phi_i \le \Phi_i^{(T=1)}.
\end{equation*}

For clarity, $\Omega_i^+(\vr)$ denotes an acceptance \emph{set} of directions; its solid-angle measure is
$|\Omega_i^+(\vr)|\equiv \iint_{\Omega_i^+(\vr)} \mathrm d\Omega$ (units \si{sr}).
Under the paraxial assumptions (B1)--(B3) (see Sec.~\ref{sec:paraxial_derivation}), the acceptance becomes approximately footprint-independent, $\Omega_i^+(\vr)\approx \Omega_i^+$,
and the cosine-weighted measure
$\mathcal G_i \equiv \iint_{\Omega_i^+}(\mathbf n_i\!\cdot\!\hat{\boldsymbol\omega})\,\mathrm d\Omega$ (units \si{sr})
appears in the reduction.

\subsection{Definition and factorization}
\label{sec:optogeometric}

The purpose of this subsection is to turn the pixel-level radiometric integral into an explicit radiance-to-flux mapping for a single detector element.
To this end, the integral is factorized into a representative (band-integrated) radiance $L_i$ and a purely geometric/throughput factor, defining the per-pixel optogeometric factor $F_{\mathrm{opg},i}$ (units \si{m^2.sr}) such that $\Phi_i \approx L_i\,F_{\mathrm{opg},i}$ under weak radiance variation.
Field- and direction-dependent pupil visibility is captured by a weighting $T_i(\mathbf r,\hat{\boldsymbol\omega})\in[0,1]$, while the ideal unvignetted baseline $T_i\equiv 1$ yields a transparent upper bound.

To obtain the compact mapping $\Phi_i \approx L_i\,F_{\mathrm{opg},i}$, the following assumptions are adopted.

\begin{enumerate}[label=(A\arabic*),ref=A\arabic*, leftmargin=*, itemsep=0.4em]

\item \textbf{Weak radiance variation:}
Within $A_{\mathrm{pix},i}\times\Omega_i^+(\mathbf r)$, radiance is approximated as constant:
$L(\mathbf r,\hat{\boldsymbol\omega})\approx L_i$ (band-integrated), with relative deviations small compared to the required accuracy.

\item \textbf{Throughput weighting:}
All field- and direction-dependent \emph{pupil visibility} effects are absorbed into
$T_i(\mathbf r,\hat{\boldsymbol\omega})\in[0,1]$.
Closed-form paraxial expressions adopt the unvignetted design baseline $T_i\equiv 1$.
\end{enumerate}

Under (A1)--(A2), the radiance is taken outside the integral, and the remaining factor directly defines the per-pixel optogeometric factor

\begin{equation}
F_{\mathrm{opg},i}
\;\equiv\;
\iint_{A_{\mathrm{pix},i}}
\iint_{\Omega_i^+(\mathbf r)}
T_i(\mathbf r,\hat{\boldsymbol\omega})\,
\bigl(\mathbf n(\mathbf r)\!\cdot\!\hat{\boldsymbol\omega}\bigr)\,
\mathrm d\Omega\,\mathrm dA,
\qquad [F_{\mathrm{opg},i}]=\si{m^2.sr}.
\label{eq:pixel_opg_factor_def}
\end{equation}

The corresponding radiance-to-flux mapping reads

\begin{equation}
\Phi_i \;\approx\; L_i\,F_{\mathrm{opg},i}.
\label{eq:pixel_opg_factor}
\end{equation}

\paragraph{Interpretation beyond (A1)}
The definition of $F_{\mathrm{opg},i}$ in Eq.~\eqref{eq:pixel_opg_factor_def} is purely geometric; assumption (A1) is only required for the factorized approximation $\Phi_i \approx L_i\,F_{\mathrm{opg},i}$.
If radiance varies across the pixel domain, one may introduce an effective band-integrated radiance as the \emph{equivalent constant radiance},
\begin{equation}
L^{\mathrm{eff}}_i \;\equiv\; \frac{\Phi_i}{F_{\mathrm{opg},i}},
\end{equation}
which is a throughput-weighted average and reduces to $L_i$ under weak radiance variation.

\subsection{Paraxial approximation of the unvignetted baseline}
\label{sec:paraxial_derivation}

This subsection derives a closed-form paraxial expression for $F_{\mathrm{opg},i}^{(T=1)}$
as a compact design baseline. The definition in Eq.~\eqref{eq:pixel_opg_factor_def} remains general and is not restricted
to the paraxial geometry considered here.

Let $\mathcal{D}_i \equiv A_{\mathrm{pix},i}\times\Omega_i^+(\mathbf r)$ denote the integration domain.
Let $\mathbf n_i$ denote the (approximately constant) unit normal over the footprint $A_{\mathrm{pix},i}$.
The following geometric assumptions are introduced for the baseline derivation:
\begin{enumerate}[label=(B\arabic*),ref=B\arabic*, leftmargin=*, itemsep=0.4em]

\item \textbf{Locally planar footprint:} the object-space surface patch associated with pixel $i$ is locally planar, so $\mathbf n(\mathbf r)\approx \mathbf n_i$ across $A_{\mathrm{pix},i}$.

\item \textbf{Weak footprint dependence of acceptance:} $\Omega_i^+(\mathbf r)\approx \Omega_i^+$ across the footprint.

\item \textbf{Common angular parametrization:} $\hat{\boldsymbol\omega}(\mathbf r;\theta,\psi)\approx \hat{\boldsymbol\omega}(\theta,\psi)$ across the footprint.

\end{enumerate}

Under (B1)--(B3) and $T_i\equiv 1$,

\begin{align}
F_{\mathrm{opg},i}^{(T=1)}
&\equiv
\iint_{A_{\mathrm{pix},i}}
\iint_{\Omega_i^+(\mathbf r)}
\bigl(\mathbf n(\mathbf r)\!\cdot\!\hat{\boldsymbol\omega}\bigr)\,
\mathrm d\Omega\,\mathrm dA
\\
&\approx
A_{\mathrm{pix},i}\,
\underbrace{\iint_{\Omega_i^+}\bigl(\mathbf n_i\!\cdot\!\hat{\boldsymbol\omega}\bigr)\,\mathrm d\Omega}_{\mathcal G_i},
\end{align}

Hence
\begin{equation}
F_{\mathrm{opg},i}^{(T=1)}\approx A_{\mathrm{pix},i}\,\mathcal G_i,
\label{eq:etendue_pixel}
\end{equation}
where $\mathcal G_i$ (units \si{sr}) is the cosine-weighted acceptance solid angle.
Under (B1)--(B3) and $T_i\equiv 1$, Eq.~\eqref{eq:etendue_pixel} therefore reduces to the pixel-level \'etendue (area--solid-angle product) in object space, consistent with \'etendue invariance in lossless passive imaging.

\subsection{Cone-specialized evaluation and design form}
\label{sec:cone_paraxial}

Assume a rotationally symmetric acceptance cone around $\mathbf n_i$ with half-angle $\alpha$:
\[
\Omega_i^+ \equiv \Omega(\alpha)=\bigl\{ \hat{\boldsymbol\omega}(\theta,\psi):\;0\le\theta\le\alpha,\;0\le\psi<2\pi\bigr\},
\quad \alpha\in\bigl(0,\tfrac{\pi}{2}\bigr].
\]
With $\mathbf n_i\!\cdot\!\hat{\boldsymbol\omega}=\cos\theta$ and $\mathrm d\Omega = \sin\theta\,\mathrm d\theta\,\mathrm d\psi$,
\begin{equation}
\mathcal G_i = \int_{0}^{2\pi}\!\!\int_{0}^{\alpha}\cos\theta\,\sin\theta\,
\mathrm d\theta\,\mathrm d\psi = \pi\,\sin^2\alpha.
\label{eq:G_cone}
\end{equation}
Thus, under (B1)--(B3),
\begin{equation}
F_{\mathrm{opg},i}^{(T=1)} \approx A_{\mathrm{pix},i}\,\pi\,\sin^2\alpha.
\label{eq:Fopg_cone}
\end{equation}

\noindent
The factor $\pi$ in Eq.~\eqref{eq:G_cone} arises from the azimuthal integration and equals the projected (cosine-weighted) solid angle of the cone; in the hemispherical limit $\alpha=\pi/2$, this projected solid angle is $\pi\,\si{sr}$.

For design-oriented reduction, the following assumptions are introduced:

\begin{enumerate}[label=(C\arabic*),ref=C\arabic*, leftmargin=*, itemsep=0.4em]

\item \textbf{Small-angle cone:} $\alpha \lesssim 0.35~\mathrm{rad}$ (e.g.\ $f\# \gtrsim 1.4$), so $\sin\alpha \approx \alpha$ (small-angle approximation).

\item \textbf{Normal incidence and square footprint:} the chief ray is near-perpendicular to the object plane.
For square detector pixels of pitch $a_{\mathrm{pp},i}$ [\si{m}], the object-space footprint area is
\[
A_{\mathrm{pix},i} = \left(\tfrac{a_{\mathrm{pp},i}}{|M|}\right)^2,
\]
where $M$ is the (signed) magnification.
\end{enumerate}

For a circular entrance pupil of diameter $D$ and focal length $f$,
\[
\tan\alpha = \frac{D}{2f}=\frac{1}{2f\#}.
\]
Combining the above yields the paraxial baseline
\begin{empheq}[box=\fbox]{equation}
F_{\mathrm{opg},i}^{(T=1)} \;\approx\; \frac{\pi}{4}\,\frac{A_{\mathrm{pix},i}}{(f\#)^2}
= \frac{\pi}{4}\,\left(\tfrac{a_{\mathrm{pp},i}}{f\#\,|M|}\right)^2 \;\equiv\; F_{\mathrm{opg},i}^{(a,f\#)} .
\label{eq:fopg_paraxial}
\end{empheq}

\subsubsection{Finite-cone correction for fast optics}
\label{sec:finite_cone_correction}

The paraxial baseline in Eq.~\eqref{eq:fopg_paraxial} relies on the small-angle approximations
$\sin\alpha\approx\alpha$ and $\tan\alpha\approx\alpha$, which are accurate when the acceptance cone is narrow.
For fast optics (small $f\#$), the cone half-angle $\alpha$ is no longer sufficiently small and the resulting $1/(f\#)^2$
throughput scaling can deviate noticeably from the exact cone value.

A simple closed-form refinement can be obtained without a full numerical evaluation of Eq.~\eqref{eq:pixel_opg_factor_def}
by (i) retaining the exact cone expression for the cosine-weighted acceptance,
$\mathcal G_i=\pi\sin^2\alpha$ from Eq.~\eqref{eq:G_cone}, and (ii) continuing to parameterize the cone half-angle
in terms of the $f$--number via the paraxial relation $\tan\alpha=1/(2f\#)$.
This illustrates a general point of the present formulation: once $F_{\mathrm{opg},i}$ is defined from the pixel-restricted radiometric integral,
alternative pixel-level closed forms follow directly by inserting an appropriate geometric model for the acceptance set (and, if needed, for pupil visibility),
rather than committing to a single canonical approximation.

Using the identity
\begin{equation}
\sin^2\alpha=\frac{\tan^2\alpha}{1+\tan^2\alpha},
\end{equation}
and the paraxial relation $\tan\alpha=1/(2f\#)$, one obtains
\begin{equation}
\sin^2\alpha
=\frac{1/(2f\#)^2}{1+1/(2f\#)^2}
=\frac{1}{4(f\#)^2+1}.
\end{equation}
Substituting this result into $F_{\mathrm{opg},i}^{(T=1)}=A_{\mathrm{pix},i}\,\pi\sin^2\alpha$ yields
\begin{equation}
F_{\mathrm{opg},i}^{(T=1)} \approx A_{\mathrm{pix},i}\,\frac{\pi}{4(f\#)^2+1}.
\label{eq:fopg_finite_cone}
\end{equation}

Equivalently, in terms of the paraxial baseline $F_{\mathrm{opg},i}^{(a,f\#)}$ defined in Eq.~\eqref{eq:fopg_paraxial},
\begin{equation}
F_{\mathrm{opg},i}^{(T=1)} \approx F_{\mathrm{opg},i}^{(a,f\#)}\;
\frac{4(f\#)^2}{4(f\#)^2+1}.
\label{eq:fopg_correction_factor}
\end{equation}
The multiplicative factor tends to unity for $f\#\gg 1$ and quantifies the departure from the small-angle baseline as $f\#$
decreases.

\section{Throughput-normalized spatial degrees of freedom and SNR implications}
\label{sec:mode_counter}

This section connects the purely geometric throughput factor $F_{\mathrm{opg},i}$ to (i) a wavelength-normalized
phase-space descriptor that is useful for comparing throughput across bands and instruments, and (ii) the resulting
photon budget and shot-noise-limited SNR scaling at the pixel level.
First, a diffraction-limited phase-space motivation is used to introduce the $\lambda^2$ normalization and the corresponding
per-pixel proxy $\mathcal{M}_i(\lambda)\equiv F_{\mathrm{opg},i}/\lambda^2$ (per polarization).
Second, the photon budget is derived directly from energy conservation and yields an explicit optics-delivered (pre-detection) SNR ceiling
whose optical dependence enters only through $F_{\mathrm{opg},i}$.

\subsection{Wavelength-normalized throughput as a phase-space proxy}
\label{sec:phase_space_proxy}

At wavelength $\lambda$ [\si{m}], diffraction-limited phase-space arguments motivate the wavelength-normalized throughput
$(A\,\Omega)/\lambda^2$ (per polarization) as a convenient proxy scaling for admitted spatial degrees of freedom
\cite{goodman2005fourier,born1999principles,welford1978high,Dandliker1999ModesETOP}.

A direct diffraction-limited motivation is that one transverse spatial degree of freedom occupies a phase-space cell of characteristic size set by the wavelength.
For example, taking a Gaussian beam as a prototypical spatial mode, the diffraction-limited divergence scales as $\theta\sim \lambda/w$,
so that a characteristic beam area $A_M\sim w^2$ and solid angle $\Omega_M\sim \theta^2$ imply $A_M\Omega_M\sim \lambda^2$
(up to convention-dependent numerical factors) \cite{Dandliker1999ModesETOP}.
This Gaussian-beam argument is used here only as an order-of-magnitude diffraction-limited motivation for the $\lambda^2$ normalization and does not enter any derivation in the paper.
Because the precise prefactors depend on adopted conventions (beam-radius and divergence definitions, full- vs.\ half-angle, truncation, etc.)
and, more broadly, on the chosen modal basis and boundary conditions, $(A\,\Omega)/\lambda^2$ is interpreted here as a geometric,
design-oriented phase-space proxy rather than a basis-independent integer ``mode count''.

In what follows, this interpretation is used only as a wavelength-normalized throughput descriptor; all quantitative SNR bounds are derived directly from the photon budget and $F_{\mathrm{opg},i}$, independent of any modal-basis interpretation.
Localizing the same throughput normalization to a single detector element then motivates the per-pixel proxy
\begin{equation}
\mathcal{M}_i(\lambda) \;\equiv\; \frac{F_{\mathrm{opg},i}}{\lambda^2},
\qquad \text{(per polarization)}, \qquad [\mathcal{M}_i]=\si{sr}.
\label{eq:Mi_def}
\end{equation}

If both polarizations are detected without polarization-selective elements, the corresponding proxy becomes $2\mathcal{M}_i$.

\subsection{Photon budget derivation}
\label{sec:photon_budget}

Photon counts follow from energy conservation.
Let the spectral radiant flux collected by pixel $i$ be
\begin{equation}
\Phi_{i,\nu}(\nu) \;=\; \tau_{\mathrm{opt}}(\nu)\,L_{\nu,i}(\nu)\,F_{\mathrm{opg},i},
\label{eq:Phi_spectral}
\end{equation}

where $L_{\nu,i}(\nu)$ is the spectral radiance seen by pixel $i$
[\si{W.m^{-2}.sr^{-1}.Hz^{-1}}], and $\tau_{\mathrm{opt}}(\nu)\in[0,1]$ [$-$] is the (dimensionless) optical transmission
(including the bandpass and other spectrally dependent throughput terms).

The role of $T_i(\mathbf r,\hat{\boldsymbol\omega})$ in Sec.~\ref{sec:radiometric} is purely geometric (pupil visibility),
whereas $\tau_{\mathrm{opt}}(\nu)$ captures spectrally dependent transmission along the optical path.

Here $\Phi_{i,\nu}(\nu)$ denotes the radiant-flux density \emph{per unit frequency} (units \si{W.Hz^{-1}}).
In the present model, $F_{\mathrm{opg},i}$ is treated as frequency-independent; any spectrally dependent throughput (bandpass, coatings, etc.) is carried by $\tau_{\mathrm{opt}}(\nu)$.
Furthermore, $L_{\nu,i}(\nu)$ denotes the object-side spectral radiance incident on the pixel acceptance set (i.e.\ prior to losses accounted for by $\tau_{\mathrm{opt}}$), so that propagation losses are not double-counted.

The detected photoelectron count accumulated over integration time $t_{\mathrm{int}}$ is

\begin{align}
N_{\mathrm{ph},i}
&=\;
t_{\mathrm{int}}\int_{\mathrm{band}} \eta(\nu)\,\frac{\Phi_{i,\nu}(\nu)}{h\nu}\,\mathrm d\nu
\\
&=\;
t_{\mathrm{int}}\int_{\mathrm{band}} \eta(\nu)\,\tau_{\mathrm{opt}}(\nu)\,
\frac{L_{\nu,i}(\nu)\,F_{\mathrm{opg},i}}{h\nu}\,\mathrm d\nu ,
\label{eq:Nph_spectral_energy}
\end{align}

where $\eta(\nu)$ is the (spectral) detector quantum efficiency.

For a narrowband channel (or weak variation across the band), at representative frequency $\bar\nu$ and bandwidth $\Delta\nu$,

\begin{align}
N_{\mathrm{ph},i}
&\;\approx\;
\eta(\bar\nu)\,t_{\mathrm{int}}\,\tau_{\mathrm{opt}}(\bar\nu)\,
\frac{L_{\nu,i}(\bar\nu)\,F_{\mathrm{opg},i}}{h\bar\nu}\,\Delta\nu
\\
&=\;
\eta(\bar\nu)\,t_{\mathrm{int}}\,\tau_{\mathrm{opt}}(\bar\nu)\,
L_{\mathrm{ph},\nu,i}(\bar\nu)\,F_{\mathrm{opg},i}\,\Delta\nu,
\label{eq:Nph_narrowband_energy}
\end{align}

where $L_{\mathrm{ph},\nu,i}(\nu)\equiv L_{\nu,i}(\nu)/(h\nu)$ is the spectral photon radiance and $\Delta\nu$ is the effective bandwidth.

At $\bar\lambda=c/\bar\nu$, substituting $F_{\mathrm{opg},i}=\mathcal{M}_i(\bar\lambda)\,\bar\lambda^2$ yields
\begin{equation}
N_{\mathrm{ph},i}
\;\approx\;
\eta(\bar\nu)\,t_{\mathrm{int}}\,\tau_{\mathrm{opt}}(\bar\nu)\,
L_{\mathrm{ph},\nu,i}(\bar\nu)\,
\bigl[\mathcal{M}_i(\bar\lambda)\,\bar\lambda^2\bigr]\,
\Delta\nu,
\label{eq:Nph_narrowband_modal}
\end{equation}
showing that the throughput-normalized proxy $\mathcal{M}_i$ provides a compact interpretation of the same energy-based photon budget.

Defining, in the narrowband approximation, the optics-delivered incident photon count as $N_{\mathrm{inc},i}\equiv N_{\mathrm{ph},i}/\eta(\bar\nu)$, the geometric scaling $N_{\mathrm{inc},i}\propto F_{\mathrm{opg},i}$ follows directly.

\subsection{SNR implications and design scalings}
\label{sec:snr}

In the photon-noise-limited regime, shot noise dominates and Poisson statistics apply \cite{HolstPerformance,Daniels2018,mandel1995}.

In photoelectron units and under Poisson statistics, shot noise has standard deviation $\sqrt{N_{\mathrm{ph},i}}$.
For a signal proportional to the detected photoelectron count, this yields the shot-noise-limited scaling
\begin{equation}
\mathrm{SNR}_i \;\approx\; \frac{N_{\mathrm{ph},i}}{\sqrt{N_{\mathrm{ph},i}}}
\;=\;\sqrt{N_{\mathrm{ph},i}}.
\label{eq:SNR_poisson}
\end{equation}

Here, $\mathrm{SNR}_i$ is expressed in photoelectron units; the corresponding optics-delivered (pre-detection) ceiling is set by the incident photon count $N_{\mathrm{inc},i}$, with $N_{\mathrm{ph},i}=\eta(\bar\nu)\,N_{\mathrm{inc},i}$ in the narrowband approximation.

Using Eq.~\eqref{eq:Nph_narrowband_energy},

\begin{align}
 \mathrm{SNR}_i
 &\;\approx\;
\sqrt{
\eta(\bar\nu)\,t_{\mathrm{int}}\,\tau_{\mathrm{opt}}(\bar\nu)\,
\frac{L_{\nu,i}(\bar\nu)\,F_{\mathrm{opg},i}}{h\bar\nu}\,\Delta\nu
}
\\
&=\;
\sqrt{
\eta(\bar\nu)\,t_{\mathrm{int}}\,\tau_{\mathrm{opt}}(\bar\nu)\,
L_{\mathrm{ph},\nu,i}(\bar\nu)\,F_{\mathrm{opg},i}\,\Delta\nu
},
\label{eq:snr_opg}
\end{align}

For a given scene and fixed detection settings (bandwidth, $t_{\mathrm{int}}$, $\tau_{\mathrm{opt}}$, and $\eta(\bar\nu)$),
Eq.~\eqref{eq:snr_opg} implies the geometric scaling
\[
\mathrm{SNR}_i \propto \sqrt{F_{\mathrm{opg},i}}.
\]

When the paraxial design baseline $F_{\mathrm{opg},i}^{(a,f\#)}$ is applicable (unvignetted baseline within the stated approximations),
the familiar dependence on pixel pitch, $f$--number, and magnification follows:
\[
\mathrm{SNR}_i \propto \frac{a_{\mathrm{pp},i}}{f\#\,|M|}.
\]
For the general weighted case, the same dependence holds with $F_{\mathrm{opg},i}$ evaluated from Eq.~\eqref{eq:pixel_opg_factor_def}
(or via an effective $\bar T_i\in[0,1]$ such that $F_{\mathrm{opg},i}=\bar T_i\,F_{\mathrm{opg},i}^{(T=1)}$).

\subsection{Bound in the presence of additional detector noise}
\label{sec:snr_bound}

When additional detector/electronics noise sources contribute \cite{HolstPerformance,Daniels2018}, the per-pixel instrument-level $\mathrm{SNR}_i$ can be expressed in photoelectron (electron-count) units as

\begin{equation}
\mathrm{SNR}_i \;=\;
\frac{N_{\mathrm{ph},i}}
{\sqrt{N_{\mathrm{ph},i} + N_{\mathrm{dark},i} + \sigma_{\mathrm{read},i}^2}}\,,
\label{eq:snr_full}
\end{equation}
where $N_{\mathrm{dark},i}$ is the expected dark-electron count over $t_{\mathrm{int}}$ and $\sigma_{\mathrm{read},i}$ is the
read-noise standard deviation (in electrons rms).
Equation~\eqref{eq:snr_full} is bounded above by the shot-noise limit, since $N_{\mathrm{dark},i}\ge 0$ and $\sigma_{\mathrm{read},i}^2\ge 0$ imply
\begin{equation}
\mathrm{SNR}_i \le \sqrt{N_{\mathrm{ph},i}} \;\propto\; \sqrt{F_{\mathrm{opg},i}}.
\end{equation}

Equivalently, in the narrowband approximation, $\mathrm{SNR}_i \le \sqrt{\eta(\bar\nu)\,N_{\mathrm{inc},i}}$. Equality holds iff $N_{\mathrm{dark},i}=0$ and $\sigma_{\mathrm{read},i}=0$.

\section{Discussion and conclusions}
\label{sec:discussion}

\subsection{Unifying interpretation}

The main practical value of the formulation is the explicit separation between pixel-level optical geometry and detector/electronics effects.
The optogeometric factor $F_{\mathrm{opg},i}$ provides a reusable geometric throughput that maps scene radiance into pixel-collected flux,
while spectrally dependent terms remain in $L_{\nu,i}(\nu)$, $\tau_{\mathrm{opt}}(\nu)$, and $\eta(\nu)$.
As a result, the pixel photon budget becomes explicit and geometry-driven $\mathrm{SNR}_i$ comparisons across aperture, vignetting,
and pixel grouping can be made without conflating them with gain/offset non-uniformity.

A key implication is that, for a given scene and fixed acquisition settings, the maximum achievable pixel $\mathrm{SNR}_i$ is constrained
already \emph{before} detection by the optics-delivered incident photon count $N_{\mathrm{inc},i}$, which scales linearly with
$F_{\mathrm{opg},i}$.
Finite quantum efficiency and detector/electronics noise sources (dark current, read noise) only map this optics-delivered budget into the achieved
electron-domain $\mathrm{SNR}_i$ and can only reduce performance below the corresponding shot-noise limit.

The throughput-normalized proxy $\mathcal{M}_i(\lambda)=F_{\mathrm{opg},i}/\lambda^2$ provides a compact wavelength normalization of pixel throughput
(per polarization) and is used here as a design-oriented spatial phase-space proxy; it is not claimed to be a basis-independent integer mode count.

\subsection{End-to-end radiance--to--SNR chain}
\label{sec:chain}

A compact way to summarize the core result is the explicit instrument-level chain that links
pixel-resolved radiometry to photon statistics:
scene radiance determines pixel-collected flux through the purely geometric throughput
$F_{\mathrm{opg},i}$; the optics-delivered photon count then follows directly, and the corresponding
shot-noise-limited SNR bound is fixed \emph{before} detection. Detector quantum efficiency and
additional noise sources only map $N_{\mathrm{inc},i}$ into the achieved electron-domain SNR and can
only reduce performance below the throughput-limited upper bound.

\begin{empheq}[box=\fbox]{align}
\text{radiance}\quad &L_i \\
\Downarrow\quad &\Phi_i \approx L_i\,F_{\mathrm{opg},i} \\
\Downarrow\quad &N_{\mathrm{inc},i}\propto F_{\mathrm{opg},i} \\
\Downarrow\quad &N_{\mathrm{ph},i}=\eta(\bar\nu)\,N_{\mathrm{inc},i} \\
\Downarrow\quad &\mathrm{SNR}_i \le \sqrt{N_{\mathrm{ph},i}}
= \sqrt{\eta(\bar\nu)\,N_{\mathrm{inc},i}}
\propto \sqrt{F_{\mathrm{opg},i}}.
\end{empheq}

\subsection{Design and calibration implications}
\label{sec:design_implications}

Because $F_{\mathrm{opg},i}$ enters the photon budget linearly, it serves as a direct design and calibration lever:
for a given scene radiance and fixed acquisition settings, improvements in geometric optical throughput translate as
$N_{\mathrm{ph},i}\propto F_{\mathrm{opg},i}$ and therefore $\mathrm{SNR}_i\propto \sqrt{F_{\mathrm{opg},i}}$ in the photon-limited regime.
This enables instrument-level trade studies to be stated at the pixel scale without mixing optical geometry with detector/electronics effects.

\textbf{Geometry-driven SNR trade studies:}
The definition~\eqref{eq:pixel_opg_factor_def} separates geometric acceptance and pupil visibility from spectral/detector terms.
As a consequence, comparisons across changes in aperture, $f$--number, vignetting, field angle, pixel pitch, or pixel binning can be made directly through the ratio
$F_{\mathrm{opg},i}^{(2)}/F_{\mathrm{opg},i}^{(1)}$.
In particular, in the paraxial unvignetted baseline,
$\mathrm{SNR}_i \propto a_{\mathrm{pp},i}/(f\#\,|M|)$ follows as a corollary, while the finite-cone refinement quantifies departures from the $1/(f\#)^2$ throughput scaling for fast optics.
This provides a compact way to estimate the SNR impact of optical design choices before resorting to detailed end-to-end simulations.

\textbf{Accounting for vignetting and field dependence:}
In practical cameras, pixel acceptance and pupil visibility vary across the field of view.
By encoding these effects through $T_i(\vr,\vomega)$, the throughput factor becomes pixel-dependent, $F_{\mathrm{opg},i}$,
and can be evaluated analytically in simplified models or numerically for instrument-specific optics.
Once a field map of $F_{\mathrm{opg},i}$ is available, its geometric contribution to spatial non-uniformity in photon counts (and thus shot-noise-limited SNR) is explicit.
This clarifies which part of observed non-uniformity is attributable to optics (throughput) rather than detector/electronics gain/offset variation, provided the calibration model does not reabsorb optical changes into the fitted gain/offset parameters.

From a phase-space viewpoint, spatial non-uniformity due to vignetting corresponds to a field-dependent truncation of the admissible ray set; this perspective has been exploited, for example, in vignetting models for light-field cameras \cite{MignardDebiseIhrke2019}.

\textbf{Implications for pixel-level calibration and NUC:}
Flat-field and non-uniformity correction procedures are typically formulated as pixel-wise gain/offset corrections applied to detector output.
The present formulation complements such procedures by making the incident radiance-to-photon mapping explicit.
In particular, gain/offset terms describe detector/electronics response non-uniformity, whereas $F_{\mathrm{opg},i}$ captures the geometric component of the radiance-to-photon budget.
This separation is useful when calibration data are interpreted across changes in aperture setting, focus, or vignetting conditions, where $F_{\mathrm{opg},i}$ may change even if detector gain/offset parameters do not.

\textbf{Throughput-normalized phase-space proxy:}
The throughput-normalized quantity $\mathcal{M}_i(\lambda)=F_{\mathrm{opg},i}/\lambda^2$ provides a convenient normalization for comparing pixel throughput across wavelength bands and instruments.
Used as a geometric proxy proportional to admitted spatial phase-space area per polarization, it supports design-level reasoning about spatial degrees of freedom without requiring a basis-dependent integer mode count.

Overall, the main practical outcome is that pixel-resolved optical throughput can be treated as a first-class quantity in instrument specifications:
it determines the optics-delivered incident-photon budget (via $N_{\mathrm{inc},i}\propto F_{\mathrm{opg},i}$) and therefore fixes the corresponding shot-noise ceiling (after including $\eta$), while sensor-level mechanisms can only reduce performance below that bound.

\textbf{Concluding remarks:}
This work elevated pixel-resolved optical throughput to a first-class geometric quantity by defining the per-pixel optogeometric factor $F_{\mathrm{opg},i}$ directly from the radiometric integral.
Because $F_{\mathrm{opg},i}$ enters the photon budget linearly, it fixes an optics-delivered shot-noise ceiling with the central scaling $\mathrm{SNR}_i \propto \sqrt{F_{\mathrm{opg},i}}$ for fixed scene and acquisition settings.
The framework remains portable across paraxial, finite-cone, vignetted, or numerical evaluations of $F_{\mathrm{opg},i}$, making geometry-driven SNR comparisons explicit and separating them cleanly from detector/electronics non-uniformity and post-processing.

\bibliographystyle{elsarticle-num}
\bibliography{references}

\end{document}